\documentclass[aps,preprint,groupedaddress]{revtex4}
\usepackage[dvips]{graphicx}
\usepackage{amsmath}
\usepackage{graphicx}
\usepackage{amsfonts}
\usepackage{amssymb}
\usepackage{epsfig}
\usepackage{subfigure}
\usepackage[usenames]{color}

\newcommand{\be}{\begin{equation}}
\newcommand{\ee}{\end{equation}}
\newcommand{\bea}{\begin{eqnarray}}
\newcommand{\eea}{\end{eqnarray}}
\newcommand{\ba}{\begin{array}}
\newcommand{\ea}{\end{array}}
\newcommand{\beas}{\begin{eqnarray*}}
\newcommand{\eeas}{\end{eqnarray*}}
\newcommand{\bes}{\begin{equation*}}
\newcommand{\ees}{\end{equation*}}

\def\i2           {\mbox{$\frac{i}{2}$}}

\begin{document}
\bibliographystyle{apsrev}

\preprint{CERN-PH-TH/2011-063}

\title{\bf Magnetic Phase Diagram of Dense Holographic Multiquarks in the Quark-gluon Plasma}

\author{Piyabut Burikham$^{1,2,3}$\thanks{Email:piyabut@gmail.com, piyabut.b@chula.ac.th}\\
{\small {\em Theoretical High-Energy Physics and Cosmology Group,
Department of Physics,}}\\
{\small {\em Faculty of Science, Chulalongkorn University, Bangkok
10330, Thailand.}}\\
$^2$ {\small {\em  Thailand Center of Excellence in Physics, CHE,
Ministry of Education, Bangkok 10400, Thailand.}}\\
$^3$ {\small {\em Department of Physics, CERN Theory Division
CH-1211 Geneva 23, Switzerland.}}\\}

\begin{abstract}
\noindent  \\
We study phase diagram of the dense holographic gauge matter in the
Sakai-Sugimoto model in the presence of the magnetic field above the
deconfinement temperature.  Even above the deconfinement, quarks
could form colour bound states through the remaining strong
interaction if the density is large.  We demonstrate that in the
presence of the magnetic field for a sufficiently large baryon
density, the multiquark-pion gradient~(MQ-$\mathcal{5}\varphi$)
phase is more thermodynamically preferred than the chiral-symmetric
quark-gluon plasma.  The phase diagrams between the holographic
multiquark and the chiral-symmetric quark-gluon plasma phase are
obtained at finite temperature and magnetic field.  In the mixed
MQ-$\mathcal{5}\varphi$ phase, the pion gradient induced by the
external magnetic field is found to be a linear response for small
and moderate field strengths.  Its population ratio decreases as the
density is raised and thus the multiquarks dominate the phase.
Temperature dependence of the baryon chemical potential, the free
energy and the linear pion gradient response of the multiquark phase
are well approximated by a simple analytic function
$\sqrt{1-\frac{T^{6}}{T^{6}_{0}}}$ inherited from the metric of the
holographic background.

\end{abstract}

\maketitle

\newpage
\section{Introduction}

Discovery of the AdS/CFT correspondence~\cite{maldacena} and the
generalization in terms of the holographic principle have provided
us with alternative theoretical methods to explore the physics of
strongly coupled gauge matter.  Holographic models have been
constructed to mimic behaviour of the strongly coupled gauge matter
in various situations.  The Sakai-Sugimoto~(SS) model~\cite{ss lowE,
ss more} is a holographic model which contains chiral fermions in
the fundamental representation of $U(N_{c})$.  Its low energy limit
is the closest holographic model of the QCD so far.  It can also
accommodate distinctively the chiral symmetry restoration and the
deconfinement phase transition in the non-antipodal case~\cite{asy}.
It provides interesting possibility of the existence of the exotic
nuclear phase where quarks and gluons are deconfined but the chiral
symmetry is still broken.

In the SS model, there are two background metrics describing a
confined and a deconfined phase.  The deconfined phase corresponds
to the background metric with a black hole horizon. The Hawking
temperature of the black hole is identified with the temperature of
the dual ``QCD" matter.  When gluons are deconfined, the
thermodynamical phase of the nuclear matter can be categorized into
3 phases, the vacuum phase, the chirally broken phase and the
chiral-symmetric phase.   In the deconfined phase, the interaction
between quarks and gluons become the screened Couloub potential. If
the coupling is still strong, bound states of quarks could form~(see
Ref.~\cite{BISY,gint,gi,cgnps,Wen,bch} for multiquark related
studies). The phase diagram of the holographic nuclear matter in the
SS model is studied in details in Ref.~\cite{bll} and extended to
include multiquarks with colour charges in Ref.~\cite{bch}.  It has
certain similarity to the conventional QCD phase diagram speculated
from other approaches e.g. the existence of critical temperature
line above which chiral symmetry is restored. The phase diagram also
shows the thermodynamic preference of the multiquark phase with
broken chiral symmetry for moderate temperature in the situation
when the density is sufficiently large. As an implication, it is
thus highly likely that matters in the core of neutron stars are
compressed into the multiquark nuclear phase. A thorough
investigation on the multiquark star suggests higher mass limits of
the neutron stars if they have multiquark cores~\cite{bhp}.

When the magnetic field is turned on, the phase structure becomes
more complicated.  Magnetic field induces the pion gradient or a
domain wall as a response of the chiral condensate of the chirally
broken phase~\cite{sst}.  In the confined phase, this is
pronounced~\cite{BLLm}.  However, it is demonstrated in
Ref.~\cite{pbm} that the pion gradient is subdominant to the
contribution from the multiquarks in the chirally broken deconfined
phase. It was also shown in Ref.~\cite{pbm} that for sufficiently
large density, the multiquark phase is more thermodynamically
preferred than the chiral-symmetric quark-gluon plasma for small
 and moderate magnetic field strengths.  Therefore it is interesting to explore
the phase diagram of the deconfined nuclear matter in the presence
of the external magnetic field.  We establish two phase diagrams
between the chirally broken multiquark~($\chi$SB) and the
chiral-symmetric quark-gluon plasma~($\chi$S-QGP), one at fixed
temperature, $T=0.10$, and another at fixed field, $B=0.20$.  The
magnetic phase diagram of the similar model for zero baryon density
is investigated in Ref.~\cite{jk}.  The phase diagram at finite
density is explored in Ref.~\cite{prs} with the approximation
$f(u)\simeq 1$.  We found that for $T\gtrsim 0.10$, this
approximation is no longer valid.

Our main results demonstrate that for a given magnetic field and
moderate temperature, the most preferred nuclear phase in the SS
holographic model is the multiquark-pion
gradient~(MQ-$\mathcal{5}\varphi$) phase provided that the density
is sufficiently large. We also study the temperature dependence of
the baryon chemical potential, the free energy, and the linear
response of the pion gradient of the mixed MQ-$\mathcal{5}\varphi$
phase and show that they inherit the temperature dependence mostly
from the SS background.

Extremely strong magnetic fields could have been produced in many
situations.  The Higgs mechanism in the cosmological electroweak
phase transition could create enormous magnetic fields in the region
between two different domains with different Higgs vacuum
expectation values~\cite{tv} which could play vital role in the
phase transitions of the nuclear soup at later times.  At the hadron
and heavy ion colliders, colliding energetic charged particles could
produce exceptional strong magnetic field locally.  The local
magnetic fields produced at RHIC and LHC are estimated to be in the
order of $10^{14-15}$ Tesla~\cite{kmw}.  On the astrophysical scale,
certain types of neutron stars called the magnetars could produce
magnetic fields as strong as $10^{10}$ Tesla~\cite{dt}.

This article is organized as the following.  In Section 2, the setup
of the deconfined SS model with additional baryon vertex and string
sources are discussed.  Main results are elaborated in Section 3.
Section 4 concludes the article.

\section{Holographic setup of the magnetized multiquark phase }

The setup we will use is the same as in Ref.~\cite{pbm}, the
Sakai-Sugomoto model with additional baryon vertex and
strings~(baryon vertex is introduced in Ref.~\cite{witb,go}).
Starting from a 10 dimensional type IIA string theory with one
dimension compactified into a circle which we will label $x^{4}$.
Two stacks of D8-branes and $\overline{\text{D8}}$-branes are then
located at distance $L$ from each other in the $x^{4}$ direction at
the boundary.  This separation will be fixed at the boundary and it
will play the role of the fundamental scale of our holographic
model. Open-string excitations with one end on the D8 and
$\overline{\text{D8}}$ will represent quarks with different
chiralities. In the background where the D8 and
$\overline{\text{D8}}$ are parallel, excitations for each chirality
are independent and there is a chiral symmetry in the background and
at the boundary.  For background with connecting D8 and
$\overline{\text{D8}}$, chiral symmetry is broken and there is a
chiral condensate.  When the energy of the connecting configuration
is minimal and there is no extra sources, we define the
corresponding boundary gauge matter to be in a vacuum phase.

Since the partition function of the string theory in the bulk is
conjectured to be equal to the partition function of the gauge
theory on the boundary, the free energy of the boundary gauge matter
is equivalent to the superstring action in the bulk~(modulo a
periodicity factor)\cite{agmoo}. We turn on non-normalizable modes
of the gauge field $a^{V}_{3},a^{A}_{1},a^{V}_{0}$~(defined in units
of $R_{D4}/2\pi \alpha'$) in the D8-branes and identify them with
the vector potential of the magnetic field, $B$~(defined in units of
$1/2\pi \alpha'$), the gradient of the chiral condensate,
$\mathcal{5}\varphi$, and the baryon chemical potential, $\mu$, at
the boundary respectively. These curious holographic correspondence
between the branes' fields and the thermodynamical quantities of the
gauge matter at the boundary allows us to study physics of the
strongly coupled non-Abelian gauge matter at finite density in the
presence of the external magnetic field.  Electric field can also be
added using other components of the gauge field on the
D8-branes~\cite{BLLm0,jk} but we will not consider such cases here.

The background spacetime of the Sakai-Sugimoto model is in the form
\begin{equation}
ds^2=\left( \frac{u}{R_{D4}}\right)^{3/2}\left( f(u) dt^2 +
\delta_{ij} dx^{i}
dx^{j}+{dx_4}^2\right)+\left(\frac{R_{D4}}{u}\right)^{3/2}\left(u^2
d\Omega_4^2 + \frac{du^2}{f(u)}\right)\\ \nonumber
\end{equation}

\begin{equation}
F_{(4)}=\frac{2\pi N}{V_4} {\epsilon}_4, \quad \quad e^{\phi}=g_s
\left( \frac{u}{R_{D4}}\right)^{3/4} ,\quad\quad R_{D4}^3\equiv \pi
g_s N_{c} l_{s}^3,\nonumber
\end{equation}

\noindent where $f(u)\equiv 1-u_{T}^{3}/u^3$, $u_T=16{\pi}^2
R_{\text{D4}}^3 {T^2} /9$.  $V_4$ is the volume of the unit
four-sphere $\Omega_4$ and $\epsilon_4$ represents the volume
4-form.  $l_{s}$ and $g_{s}$ are the string length scale and the
string coupling respectively.  $R$ is the compactified radius of the
$x^{4}$ coordinate.  This radius is different from the curvature
$R_{D4}$ of the background in general.  The dilaton field is denoted
by $\phi$ which will be eliminated by the function of $u$ as stated
above.

The direction of the magnetic field is chosen so that the vector
potential is
\begin{eqnarray}
a^{V}_{3}& = & B x_{2}.
\end{eqnarray}
The baryon chemical potential $\mu$ of the corresponding gauge
matter is identified with the non-normalizable mode of the DBI gauge
field at the boundary by
\begin{eqnarray}
\mu & = & a^{V}_{0}(u\to\infty).
\end{eqnarray}

The five-dimensional Chern-Simon term of the D8-branes generates
another axial part of the $U(1)$, $a^{A}_{1}$, by coupling it with
$B$ and $a^{V}_{0}$.  In this way, the external magnetic field
induces the axial current $j_{A}$ associated with the axial field
$a^{A}_{1}$.  The non-normalizable mode of this field at the
boundary corresponds to the response of the chiral condensate to the
magnetic field which we call the pion gradient,
$\mathcal{5}\varphi$.  External field causes the condensate to form
a domain wall which can be characterized by the gradient of the
condensate with respect to the direction of the applied field.
Therefore the pion gradient also acts as a source of the baryon
density in our gauge matter.

Additional sources of the baryon density and the baryon chemical
potential can be added to the configuration in the form of the
baryon vertex and strings~\cite{bll, bch}.
\begin{eqnarray}
S_{source}     & = & {\mathcal N} d \Big[
\frac{1}{3}u_{c}\sqrt{f(u_{c})}+n_{s}(u_{c}-u_{T})\Big],
\\ \label{app4}
               & = & {\mathcal N} d \mu_{source}
\end{eqnarray}
where $n_{s}=k_{r}/N_{c}$ is the number of radial strings in the
unit of $1/N_{c}$.  Since the radial strings could merge with
strings from other multiquark and generate a binding potential
between the multiquarks, this number therefore represents the colour
charges of the multiquark in the deconfined phase.  It is
interesting to note that when there is only string source
representing quark matter, the quark matter becomes
thermodynamically unstable under density fluctuations~\cite{bll}.
However, adding baryon vertex together with the strings makes the
multiquark configuration stable under the density
fluctuations~\cite{bch}.  The multiquark phase is even more
thermodynamically preferred than the $\chi$S-QGP when the density is
sufficiently large and the temperature is not too high.

With this setup, then the DBI and the Chern-Simon actions of the
D8-branes configuration can be calculated to be
\begin{eqnarray}
S_{D8}& = & \mathcal{N}
\int^{\infty}_{u_{c}}du~u^{5/2}\sqrt{1+\frac{B^{2}}{u^{3}}}\sqrt{1+f(u)(a_{1}^{\prime
A})^{2}-(a_{0}^{\prime V})^{2}+f(u)u^{3}x_{4}^{\prime 2}},  \label{action1}  \\
S_{CS}& = & -\frac{3}{2}\mathcal{N}
\int^{\infty}_{u_{c}}du~(\partial_{2}a^{V}_{3}a^{V}_{0}a^{A
\prime}_{1}-\partial_{2}a^{V}_{3}a^{V \prime}_{0}a^{A}_{1}),
\label{action2}
\end{eqnarray}
where $\mathcal{N}=NR^{2}_{D4}/(6\pi^2(2\pi \alpha^{\prime})^{3})$
defines the brane tension.  The factor $3/2$ in the Chern-Simon
action fixes the edge effect of the region with uniform magnetic
field as explained in Ref.~\cite{BLLm}.

We can write down the equations of motion with respect to each gauge
field $a_{0}^{V},a_{1}^{A}$ as
\begin{eqnarray}
\frac{\sqrt{u^{5}+B^{2}u^{2}}~f(u)a_{1}^{\prime
A}}{\sqrt{1+f(u)(a_{1}^{\prime A})^{2}-(a_{0}^{\prime
V})^{2}+f(u)u^{3}x_{4}^{\prime 2}}}& = & j_{A}-\frac{3}{2}B\mu+3B
a_{0}^{V}, \label{eq:a0} \\
\frac{\sqrt{u^{5}+B^{2}u^{2}}~a_{0}^{\prime
V}}{\sqrt{1+f(u)(a_{1}^{\prime A})^{2}-(a_{0}^{\prime
V})^{2}+f(u)u^{3}x_{4}^{\prime 2}}}& = &
d-\frac{3}{2}Ba_{1}^{A}(\infty)+3B a_{1}^{A}. \label{eq:a1}
\end{eqnarray}
$d,j_{A}$ are the corresponding density and current density at the
boundary of the background~($u\to\infty$) given by
\begin{eqnarray}
j^{\mu}(x, u\to\infty)& \equiv & \frac{\delta S_{eom}}{\delta
A_{\mu}}\bigg{\vert}_{u\to\infty} \\
                      & \equiv & (d,\vec{j_{A}}).
\end{eqnarray}
In terms of the gauge fields, they are
\begin{eqnarray}
d & = & \frac{\sqrt{u^{5}+B^{2}u^{2}}~a_{0}^{\prime
V}}{\sqrt{1+f(u)(a_{1}^{\prime A})^{2}-(a_{0}^{\prime
V})^{2}+f(u)u^{3}x_{4}^{\prime 2}}}\bigg{\vert}_{\infty}-\frac{3}{2}B a_{1}^{A}(\infty), \\
j_{A}& = & \frac{\sqrt{u^{5}+B^{2}u^{2}}~f(u)a_{1}^{\prime
A}}{\sqrt{1+f(u)(a_{1}^{\prime A})^{2}-(a_{0}^{\prime
V})^{2}+f(u)u^{3}x_{4}^{\prime
2}}}\bigg{\vert}_{\infty}-\frac{3}{2}B\mu.
\end{eqnarray}
In order to solve these equations, we need to specify the boundary
conditions.  Due to the holographic nature of the background
spacetime, the boundary conditions correspond to physical
requirement we impose to the gauge matter. If we want to address
chirally broken phase of the gauge matter, we will take
$a^{A}_{1}(\infty)\equiv \mathcal{5}\varphi$ to be an order
parameter of the chiral symmetry breaking and minimize the action
with respect to it.  This results in setting $j_{A}=0$. On the other
hand, if we want to study the chiral-symmetric gauge matter~(or
chiral-symmetric quark-gluon plasma for $N_{c}=3$ case), $x'_{4}$
and $a^{A}_{1}(\infty)$ will be set to zero.  For vacuum phase,
$a_{0}^{V}, a_{1}^{A}$ and $d, j_{A}$ will be set to zero.

In any cases, since the total action does not depend on $x_{4}(u)$
explicitly, the constant of motion gives
\begin{eqnarray}
(x^{\prime}_{4}(u))^{2}& = & \frac{1}{u^{3}f(u)}\Big[
\frac{u^{3}[f(u)(C(u)+D(u)^{2})-(j_{A}-\frac{3}{2}B\mu
+3Ba_{0}^{V})^{2}]}{F^{2}}-1 \Big]^{-1},  \label{eq:x4prime}
\end{eqnarray}
where
\begin{eqnarray}
F & = & \frac{u^{3}_{c}
\sqrt{f(u_{c})}\sqrt{f(u_{c})(C(u_{c})+D(u_{c})^{2})-(j_{A}-\frac{3}{2}B\mu
+3Ba_{0}^{V}(u_{c}))^{2}}~x_{4}^{\prime}(u_{c})}{\sqrt{1+f(u_{c})u^{3}_{c}~x_{4}^{\prime
2}(u_{c})}}  \label{eq:F}
\end{eqnarray}
and $C(u)\equiv u^{5}+B^{2}u^{2},D(u)\equiv
d+3Ba_{1}^{A}(u)-3B\mathcal{5}\varphi/2$. The calculation of
$x_{4}^{\prime}(u_{c})$ is described in the Appendix as a result
from the equilibrium and scale fixing condition
\begin{equation}
L_{0} = 2 \int^{\infty}_{u_{c}}x_{4}^{\prime}(u)~du = 1.
\label{eq:sf}
\end{equation}
The equations of motion Eqn.~(\ref{eq:a0}),(\ref{eq:a1}) can be
solved numerically under the constraint (\ref{eq:sf}).  The value of
$\mu, \mathcal{5}\varphi, u_{c}$ and the initial values of
$a^{V}_{0}(u_{c}),a^{A}_{1}(u_{c})$ are chosen so that
$a^{V}_{0}(\infty)=\mu,a^{A}_{1}(\infty)=\mathcal{5}\varphi$ and
$L_{0}=1$ are satisfied simultaneously.

\section{Magnetic phase diagram of the dense nuclear phase}

Generically, the action (\ref{action1}) and (\ref{action2}) are
divergent from the $u\to \infty$ limit of the integration and we
need to regulate it using the action of the vacuum which is also
divergent. The contribution from the region $u\to \infty$ is
divergent even when the magnetic field is turned off and it is
intrinsic to the DBI action in this background. The divergence can
be understood as the infinite zero-point energy of the system and
thus could be systematically removed by regularisation.

Therefore the regulated free energy is given by
\begin{eqnarray}
\mathcal{F}_{\text{E}}=\Omega(\mu,B) +\mu d,
\end{eqnarray}
where $\Omega(\mu,B) =
S[a_{0}(u),a_{1}(u)](e.o.m.)-S[\text{magnetized vacuum}]$.  Note
that we need to Legendre transform the DBI and the Chern-Simon
action to obtain the bulk action as a function of the
non-normalizable modes $a^{V}_{0}, a^{A}_{1}$ in order to identify
it with the free energy of the gauge matter at the boundary.  In
terms of the free energy at the boundary, this is equivalent to the
Legendre transform of the grand canonical with respect to $\mu$ and
$d$.

We can calculate the total action satisfying the equation of motion
$S[a_{0}(u),a_{1}(u)](e.o.m.)= S_{D8}+S_{CS}$ to be
\begin{eqnarray}
S_{D8}& = &\mathcal{N}
\int^{\infty}_{u_{c}}du~C(u)\sqrt{\frac{f(u)(1+f(u)u^{3}{x^{\prime
2}_{4})}}{f(u)(C(u)+D(u)^{2})-(j_{A}-\frac{3}{2}B\mu
+3Ba_{0}^{V})^{2}}}, \\
S_{CS}& = & -\mathcal{N} \frac{3}{2}B
\int^{\infty}_{u_{c}}du~\frac{\Big( a^{V}_{0}(j_{A}-\frac{3}{2}B\mu
+3Ba_{0}^{V})-f(u)
D(u)a^{A}_{1}\Big)\sqrt{\frac{1}{f(u)}+u^{3}x_{4}^{\prime
2}}}{\sqrt{f(u)(C(u)+D(u)^{2})-\Big(j_{A}-\frac{3}{2}B\mu
+3Ba_{0}^{V}\Big)^{2}}}. \label{CSL}
\end{eqnarray}

The three nuclear phases above the deconfinement temperature are
governed by the same equations of motion, each with specific
boundary conditions as the following, \\

\underline{magnetized vacuum phase}: $a^{V}_{0}, a^{A}_{1}=0;
d, j_{A}=0,$  \\

\underline{multiquark-pion gradient phase}: $a^{V}_{0}(u_{c})=
\mu_{source}, a^{A}_{1}(u_{c})=0, a^{A}_{1}(\infty)=
\mathcal{5}\varphi, j_{A}=0,$ \\

\underline{$\chi$S-QGP phase}: $x'_{4}(u)=0,
a^{V}_{0}(u_{c}=u_{T})=0, a^{A}_{1}(\infty)=0, j_{A}=
\frac{3}{2}B\mu.$ \\

We will demonstrate later that in the mixed phase, the pion gradient
is generically dominated by the multiquark when the chiral symmetry
is broken.  In Ref.~\cite{pbm}, it is shown that the pure pion
gradient phase is always less preferred thermodynamically than the
mixed phase of MQ-$\mathcal{5}\varphi$. It is interesting to note
that for the pure pion gradient phase, a large magnetic field is
required in order to stabilize the generated domain wall~\cite{sst}.
This critical field is determined by the mass of the pion in the
condensate, $B_{crit}\sim m^{2}_{\pi}/e$.  In Ref.~\cite{pbm}, this
critical behaviour is confirmed in the holographic SS model~(the
zero-temperature situation is studied in Ref.~\cite{ts}). More
investigation of the pure pion gradient phase in the holographic
model should be conducted especially when the field is large since
the distinctive feature of physics from the DBI action becomes
apparent in this limit.  We will leave this task for future work and
focus our attention to the mixed MQ-$\mathcal{5}\varphi$ phase in
this article.

\begin{figure}[htp]
\centering
\includegraphics[width=0.6 \textwidth]{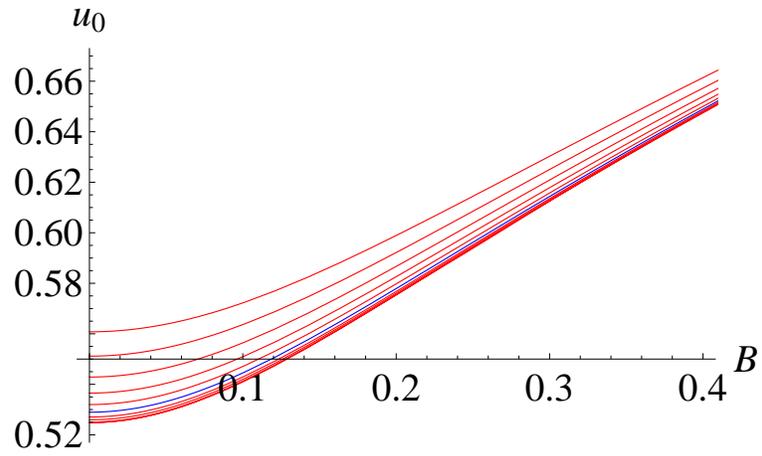}
\caption[u0]{ The position $u_{0}$ as a function of $B$ of the
magnetized vacuum for $T = 0.02-0.15$.  The upper lines have higher
temperatures. } \label{fig1}
\end{figure}

The action of the magnetized vacuum when we set $a^{V}_{0},
a^{A}_{1}=0$ and $d, j_{A}=0$ is
\begin{eqnarray}
S[\text{magnetized vacuum}] & = &
\int^{\infty}_{u_{0}}~\sqrt{C(u)(1+f(u)u^{3}x^{\prime
2}_{4})}\bigg{\vert}_{vac}~du, \nonumber
\end{eqnarray}
where
\begin{eqnarray}
 x^{\prime}_{4}(u)\vert_{vac} & = &
\frac{1}{\sqrt{f(u)u^{3}\Big(\frac{f(u)u^{3}C(u)}{f(u_{0})u^{3}_{0}C(u_{0})}-1
\Big)}}.
\end{eqnarray}
The position $u_{0}$ where $x'_{4}\to \infty$ of the magnetized
vacuum configuration increases slightly with temperature as is shown
in Fig.~\ref{fig1}.  The difference between each temperature
decreases as the magnetic field gets larger and all curves converge
to the same saturated value $u_{0}=1.23$ in the large field limit.

\begin{figure}[htp]
\centering
\includegraphics[width=0.45\textwidth]{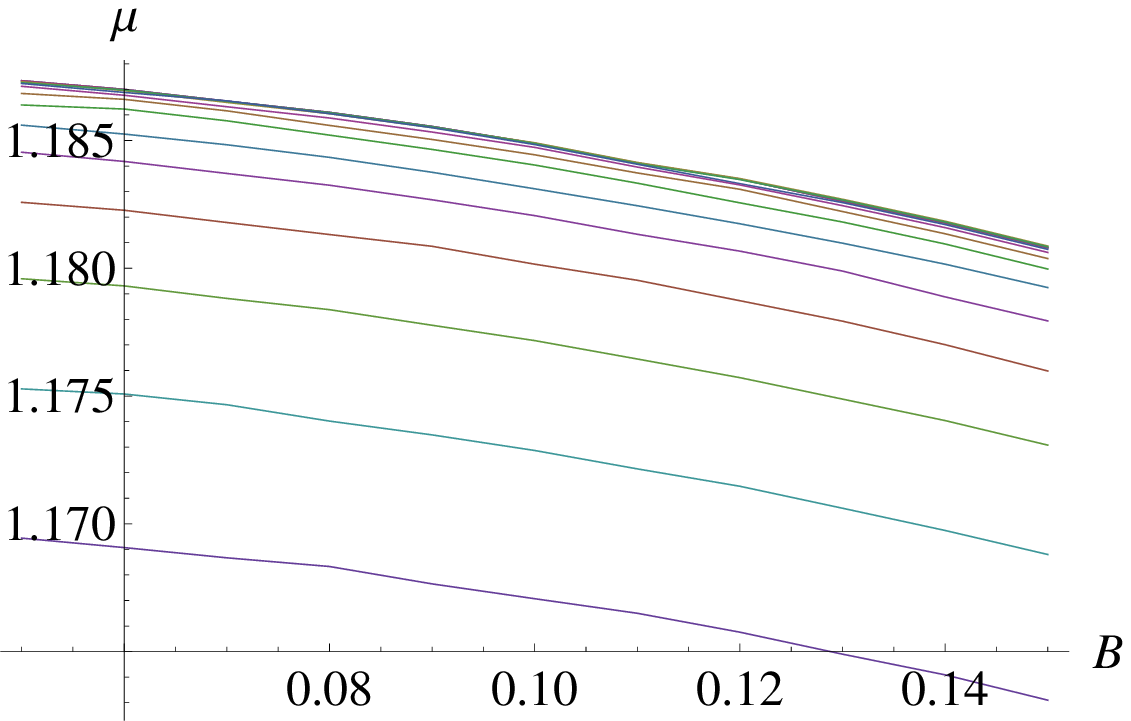} \hfill
\includegraphics[width=0.45\textwidth]{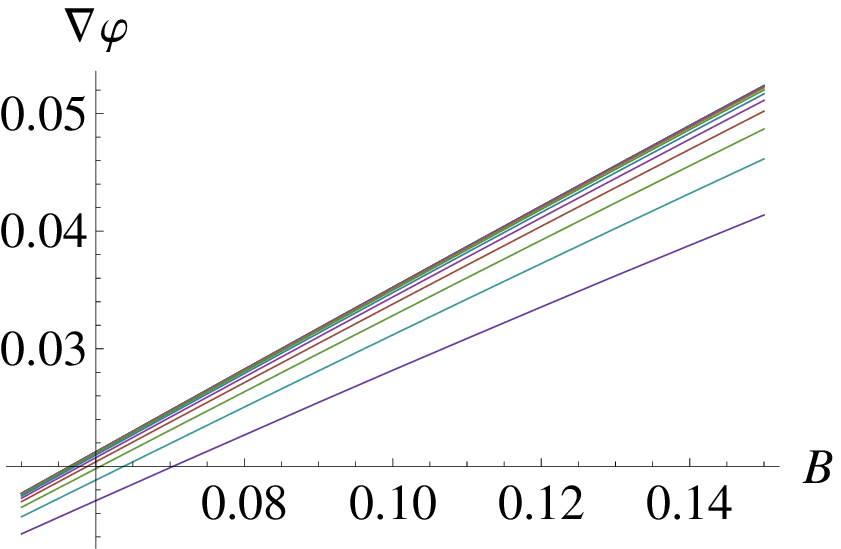}\\
\includegraphics[width=0.45\textwidth]{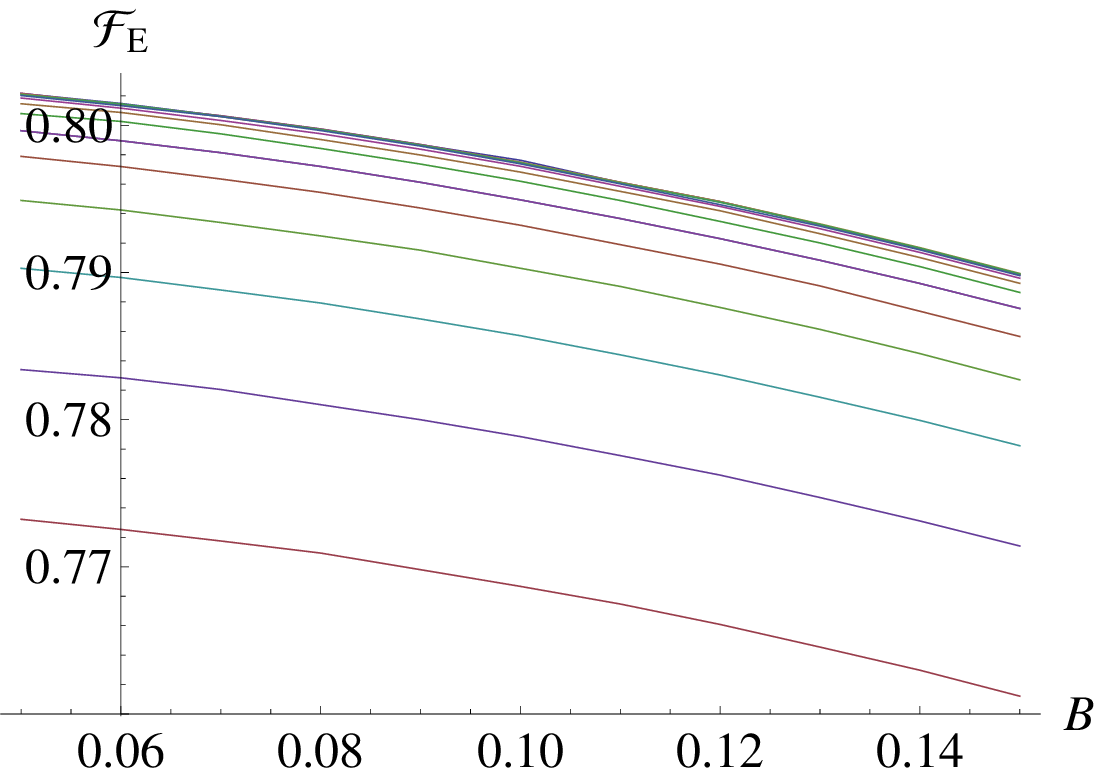}\\
\caption[TBd1]{ The chemical potential~(a), the pion gradient~(b)
and the free energy~(c) of the multiquark phase with baryon density
$d=1$ as a function of $B$ for temperature $T=0.02-0.15$.  The lower
curves represent multiquark at higher temperatures. } \label{fig1.0}
\end{figure}

\begin{figure}[htp]
\centering
\includegraphics[width=0.45\textwidth]{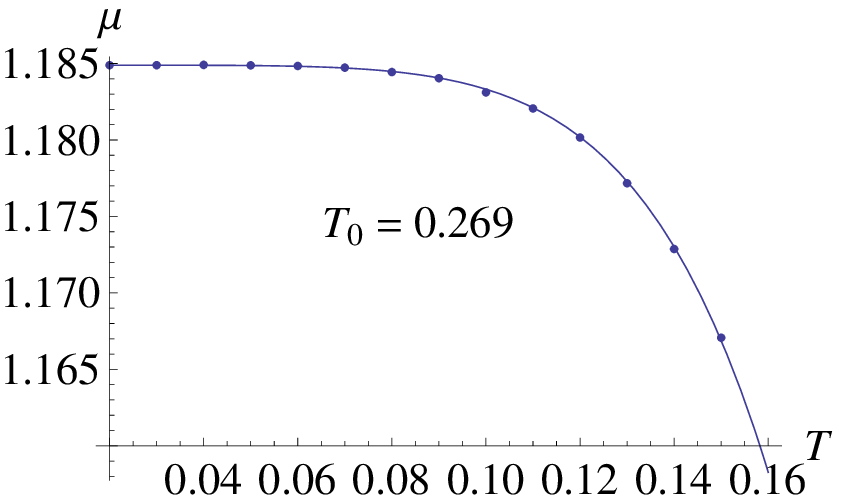} \hfill
\includegraphics[width=0.45\textwidth]{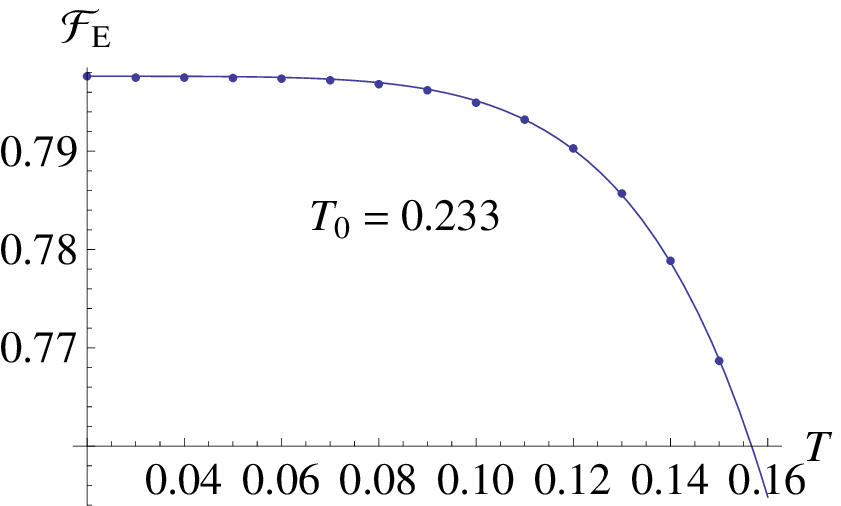}
\caption[muFreeT]{ For $d=1, B=0.10$,(a) the baryon chemical
potential as a function of $T$, the best-fit curve is in the form
$\mu_{0}\sqrt{1-(\frac{T}{T_{0}})^{6}}$ with $\mu_{0}=1.1849,
T_{0}=0.269$; (b) the free energy as a function of $T$, the best-fit
curve is in the form $F_{0}\sqrt{1-(\frac{T}{T_{0}})^{6}}$ with
$F_{0}=0.7976, T_{0}=0.233$.  Other curves within the range
$B=0.05-0.15$ can also be fitted well with the same $T_{0}.$ }
\label{fig1.1b}
\end{figure}

We can study the temperature dependence of the magnetized multiquark
nuclear matter by considering its baryon chemical potential and the
free energy as shown in Fig.~\ref{fig1.0}.  Both the chemical
potential and the free energy decrease steadily as the temperature
rises, regardless of the magnetic field.  This is originated from
the temperature dependence of $f(u)=1-\frac{u_{T}^{3}}{u^{3}}$ of
the SS background in the deconfined phase.  The temperature
dependence could be fit very closely with the function
$\sqrt{1-(\frac{T}{T_{0}})^{6}}$ as the following
\begin{eqnarray}
\mu & = & \mu_{0}(d,B)\sqrt{1-(\frac{T}{T_{0}})^{6}}, \\
F  & = & F_{0}(d,B)\sqrt{1-(\frac{T}{T_{0}})^{6}}.
\end{eqnarray}
where for $d=1,B=0.10$; $\mu_{0}=1.1849, F_{0}=0.7976$ respectively.
For the baryon chemical potential~(free energy), the best-fit value
of $T_{0}$ is $0.269~(0.233)$. The fittings are shown in
Fig.~\ref{fig1.1b}. This could be explained by noting that the
regulated free energy is given by $\mu d + \Omega(\mu,B)$. The
contribution from the first term is dominant therefore the free
energy has almost the same temperature dependence as the chemical
potential.  However, there is a minor contribution from
$S_{\text{D8}}+S_{\text{CS}}$ containing
$f(u_{c})=1-\frac{u_{T}^{3}}{u_{c}^{3}}$ which for small temperature
fractions modifies the temperature function in the following manner,
\begin{eqnarray}
C_{1}\sqrt{1-\frac{T^{6}}{T^{6}_{1}}} +
C_{2}\sqrt{1-\frac{T^{6}}{T^{6}_{2}}} & \simeq & C_{0}
\sqrt{1-\frac{T^{6}}{T^{6}_{0}}},
\end{eqnarray}
where $C_{1,2}$ are some arbitrary constants and $C_{0},T_{0}$ are
given by
\begin{eqnarray}
C_{0} & = & C_{1}+C_{2},  \\
\frac{1}{T^{6}_{0}} & = &
\frac{1}{C_{1}+C_{2}}\left(\frac{C_{1}}{T^{6}_{1}} +
\frac{C_{2}}{T^{6}_{2}}\right).
\end{eqnarray}
It should be noted from Fig.~\ref{fig1.1b} that the temperature
dependence is significant for $T \gtrsim 0.10$ and the approximation
$f(u)\simeq 1$ is not accurate for temperature in this range.  The
characteristic temperatures we found here are consistent with the
phase diagram of the multiquark in Fig.~\ref{fig1.3}.

In the multiquark phase when the magnetic field is turned on, the
pion gradient is induced by the field in addition to the multiquark.
The multiquark phase thus contained the mixed content of multiquarks
and the pion gradient.  For moderate fields~(not too large), the
response is linear $\mathcal{5}\varphi \propto B$.  In contrast to
the case of pure pion gradient phase, the domain wall in the mixed
MQ-$\mathcal{5}\varphi$ phase is stable among the surrounding
multiquarks even for small field.  The critical magnetic field to
stabilize the domain wall in the case of pure pion gradient is not
required in the mixed phase.

\begin{figure}[htp]
\centering
\includegraphics[width=0.65\textwidth]{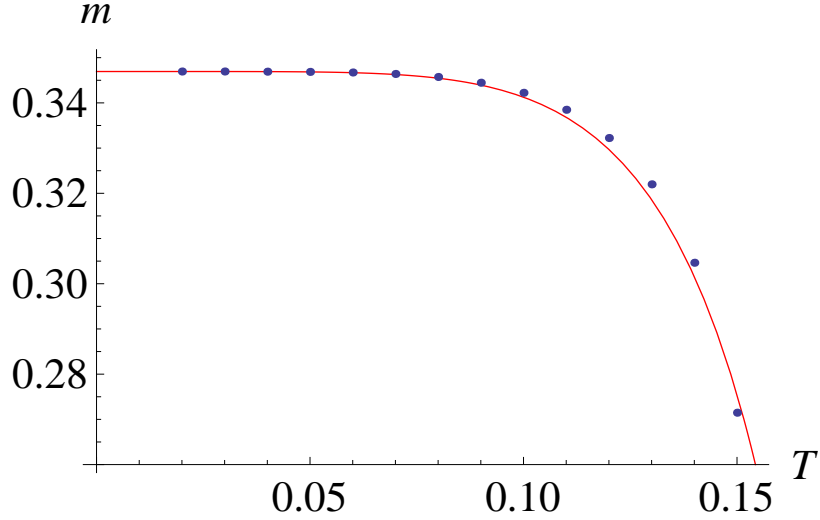}
\caption[slopeyBT]{The linear response or slope of the linear
function between the pion gradient and the magnetic field as a
function of the temperature for the range $B = 0.05-0.15$ and
density $d=1$. The red line is the best-fit curve in the form
$m_{0}\sqrt{1-(\frac{T}{T_{0}})^{6}}$ with $m_{0}=0.347,
T_{0}=0.177$. } \label{fig1.1}
\end{figure}

\begin{figure}[htp]
\centering
\includegraphics[width=0.45\textwidth]{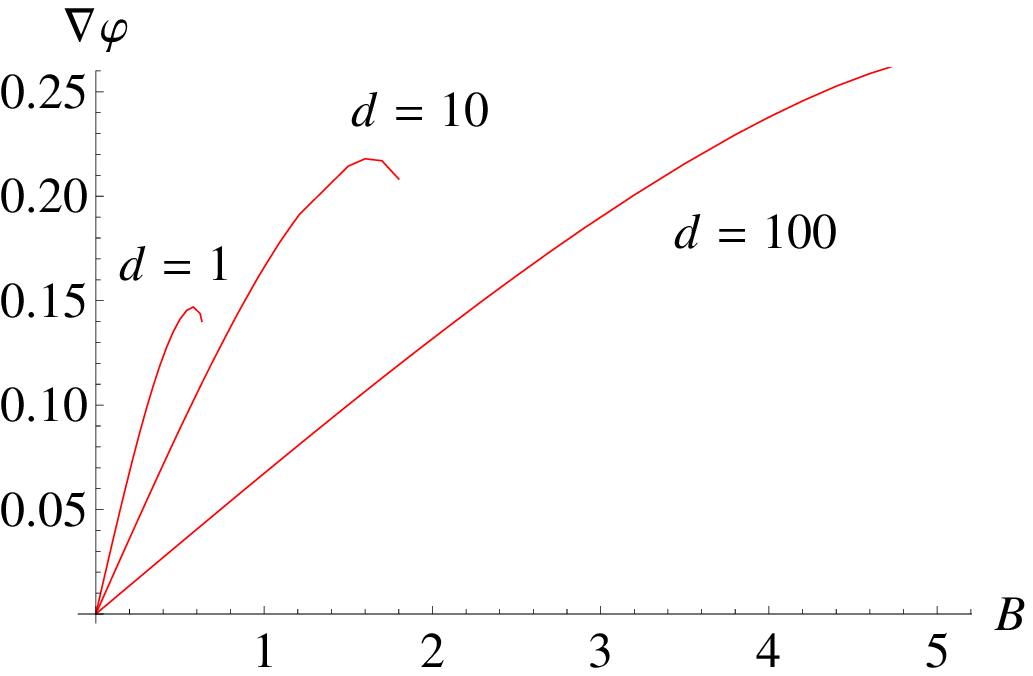} \hfill
\includegraphics[width=0.45\textwidth]{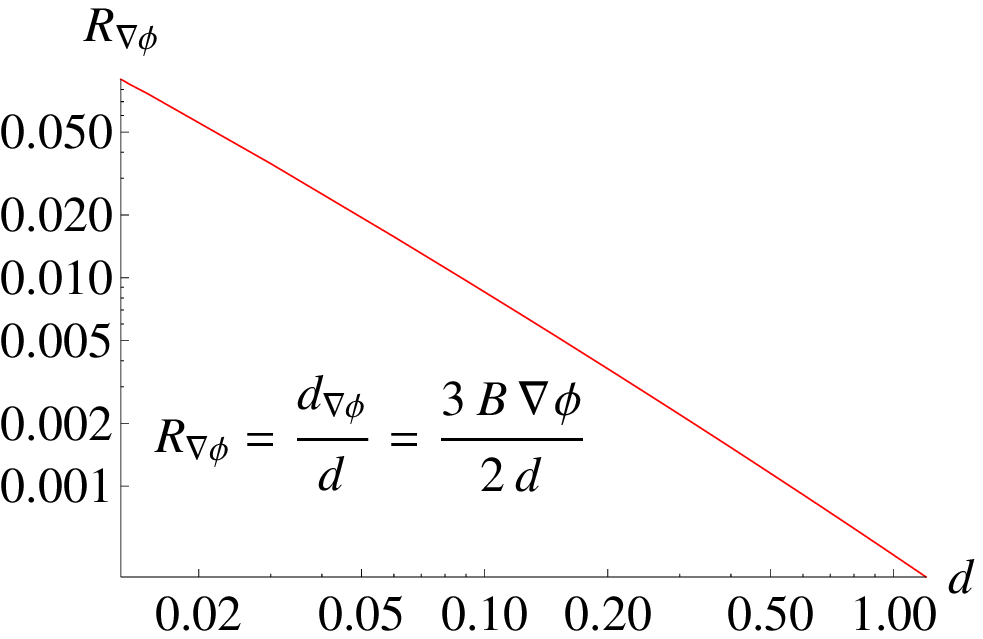}
\caption[slopeyBT]{(a) The pion gradient as a function of $B$ for
density $d = 1, 10, 100$ at $T = 0.10$.  (b) The density ratio of
the pion gradient with respect to the total baryon density of the
multiquark phase at $B=0.10, T=0.10$ in the double-log scale. }
\label{fig1.11}
\end{figure}

Figure \ref{fig1.0}~(b) shows a linear relation between the pion
gradient and the magnetic field which is valid up to moderate
fields.  For $d=1$, we found that the slope, $m$~(or the linear
response), of this linear function depends on the temperature
approximately as $m = m_{0}\sqrt{1-(\frac{T}{T_{0}})^{6}}$, and
\begin{eqnarray}
\mathcal{5}\varphi & \simeq & B m_{0}
\sqrt{1-\left(\frac{T}{T_{0}}\right)^{6}},  \label{piong}
\end{eqnarray}
where $m_{0}=0.347, T_{0}=0.177$.  The curve fitting is shown in
Fig.~\ref{fig1.1}. The density dependence is encoded in
$m_{0}=m_{0}(d), T_{0}=T_{0}(d)$. As the density increases, the
slope of the linear response of the pion gradient becomes smaller as
is shown in Fig.~\ref{fig1.11}. The ratio of the pion gradient
density and the total baryonic density $R_{\mathcal{5}\varphi}\equiv
d_{\mathcal{5}\varphi}/d = 3B\mathcal{5}\varphi/2d$~\cite{BLLm} for
$B=0.10, T=0.10$ is plotted in the log-scale in
Fig.~\ref{fig1.11}~(b).  It could be well approximated by
\begin{eqnarray}
R_{\mathcal{5}\varphi}& \simeq &(\text{const.}) d^{-6/5}, \\
& \simeq &
\frac{3B^{2}m_{0}}{2d}\sqrt{1-\left(\frac{T}{T_{0}}\right)^{6}},
\end{eqnarray}
from Eqn.~(\ref{piong}).  This implies that the multiquark states
are more preferred than the pion gradient in the presence of the
magnetic field, the denser the nuclear matter, the more stable the
multiquarks become and the lesser the population of the pion
gradient.

\begin{figure}[htp]
\centering
\includegraphics[width=0.45\textwidth]{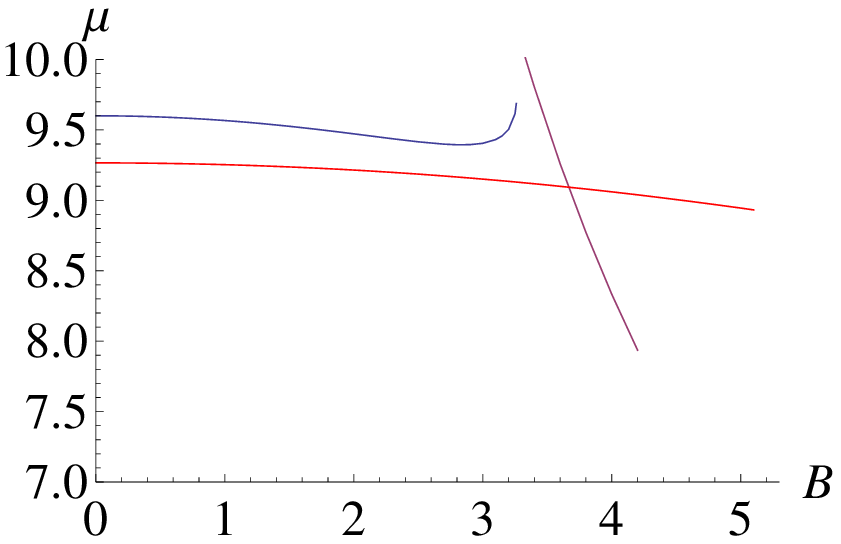} \hfill
\includegraphics[width=0.45\textwidth]{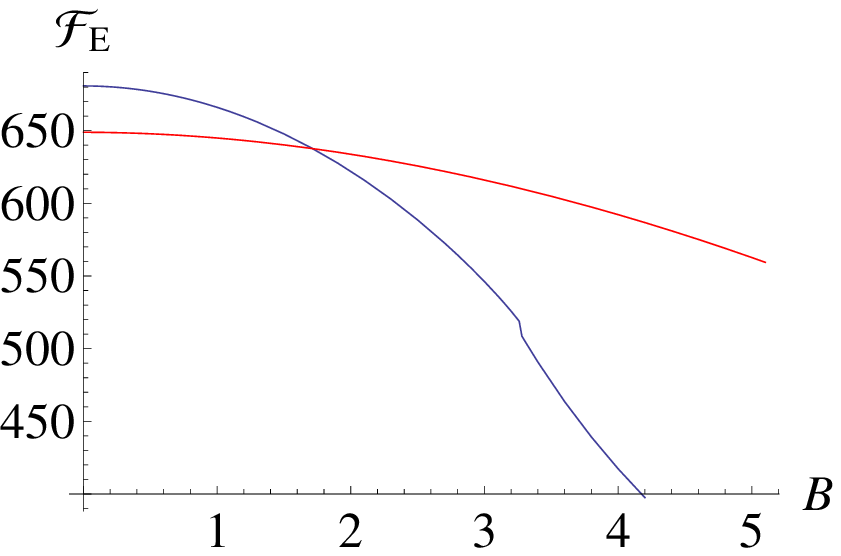}\\
\caption[mqd100]{ For the dense multiquark with $d=100, T = 0.10$,
(a) the chemical potential, (b) the free energy as a function of
$B$.  The multiquark curves in red are compared with the $\chi$S-QGP
curves in blue for the chemical potential and the free energy. }
\label{fig1.2}
\end{figure}

\begin{figure}[htp]
\centering
\includegraphics[width=0.45\textwidth]{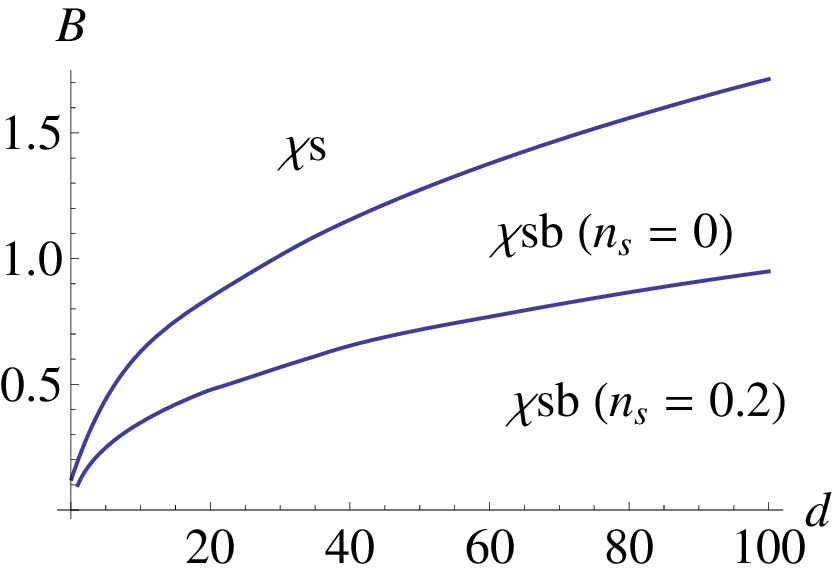} \hfill
\includegraphics[width=0.45\textwidth]{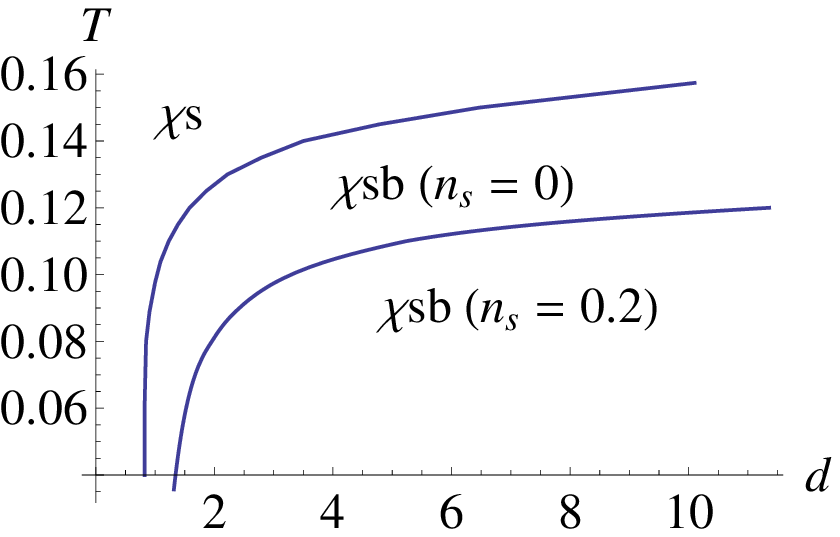}\\
\caption[phase diagram]{The phase diagram of the dense nuclear
phases involving multiquarks when gluons are deconfined for (a) $T =
0.10$ and (b) $B = 0.20$.  The chiral-symmetric quark-gluon plasma
and the chirallly broken MQ-$\mathcal{5}\varphi$ phase are
represented by $\chi$S and $\chi$SB respectively, $n_{s}$ is the
number of colour strings in fractions of $1/N_{c}$. } \label{fig1.3}
\end{figure}

Finally we compare the free energy of the MQ-$\mathcal{5}\varphi$
phase and the chiral-symmetric quark-gluon plasma phase. For high
density, $d=100$, this is shown in Fig.~\ref{fig1.2}.  For a given
density, the multiquark phase is more thermodynamically preferred
than the $\chi$S-QGP for small and moderate fields.  As the magnetic
field gets larger, the $\chi$S-QGP becomes more thermodynamically
preferred. When the field becomes very strong, the transition into
the lowest Landau level finally occurs~\cite{ll}.  For a fixed
density, increasing magnetic field inevitably results in the chiral
symmetry restoration.  The phase transition between the
MQ-$\mathcal{5}\varphi$ and the $\chi$S-QGP is a first order since
the free energy is continuous at the transition and the slope has a
discontinuity.  It implies that the magnetization,
$M(d,B)=-\frac{\partial \mathcal{F}_{\text{E}} }{\partial B}$, of
the nuclear matter abruptly changes at the transition.

On the other hand, for a fixed field and the moderate temperature,
the increase in the baryon density could make the multiquark phase
more stable than the $\chi$S-QGP.  This is shown in the phase
diagram in Fig.~\ref{fig1.3}. At a given magnetic field, the
multiquark phase could become the most preferred magnetized nuclear
phase provided that the density is made sufficiently large and the
temperature is not too high.  In contrast, the effect of the
temperature is the most dominant for chiral-symmetry restoration
even when the field is turned on.  Sufficiently large temperature
will induce chiral-symmetry restoration for most densities as is
shown the Fig.~\ref{fig1.3}(b).  Nevertheless, theoretically we can
always find sufficiently large density above which the multiquark
phase is more preferred.

The transition line between the MQ-$\mathcal{5}\varphi$ and the
$\chi$S-QGP phases in the $(d,B)$ phase diagram can be approximated
by a power-law
\begin{eqnarray}
B & \sim & d^{0.438~(0.436)}
\end{eqnarray}
for the multiquark with $n_{s}=0~(0.2)$. This power-law is weaker
than the transition line of the $\chi$S-QGP to the lowest Landau
level studied in Ref.~\cite{ll} for the antipodal SS model~($B \sim
d^{2/3}$).  The multiquarks with more colour charges~($n_{s}>0$) are
less preferred thermodynamically but they require higher densities.
On the other hand, the transition line in the $(d,T)$ phase diagram
is an increasing function of $d$ but weaker than the logarithmic of
the density. Nevertheless, theoretically for a fixed $B,T$, we can
always find sufficiently large density above which the
MQ-$\mathcal{5}\varphi$ phase is preferred.  The high density region
is actually dominated by the multiquark phase indeed.

\section{Conclusion}

We explore the properties of the miltiquark-domain
wall~(MQ-$\mathcal{5}\varphi$) solution of the SS model above the
deconfinement temperature.  The temperature dependence of the baryon
chemical potential, the pion gradient linear response~($m$), and the
free energy of the MQ-$\mathcal{5}\varphi$ phase has been studied
and fitted with a simple function,
$\sqrt{1-\frac{T^{6}}{T^{6}_{0}}}$, inherited from the deconfined SS
background.  Their characteristic temperatures, $T_{0}$, are
different from one another depending on other parameters such as
$u_{c}$, the position of the baryon vertex.  Remarkably, they do not
depend on the field for moderate field strength $B=0.05-0.15$.

For chirally broken deconfined nuclear matter in the presence of the
magnetic field, the nuclear matter with finite baryon density and
chemical potential could respond to the magnetic field by inducing a
pion gradient or a domain wall of the chiral condensate.  This pion
gradient response is found to be a linear function of the field for
moderate fields at any density.  However, we demonstrate further
that the population ratio of the pion gradient decreases as the
density increases.  The other sources of the baryon charge namely
the multiquarks finally dominate the chirally broken nuclear phase
and most of the baryon density is in the form of the multiquark at
high density.

Magnetic phase diagram of the dense gauge matter have been explored
in the deconfined SS model.  At fixed magnetic field and moderate
temperature, the MQ-$\mathcal{5}\varphi$ phase are more preferred
than the $\chi$S-QGP for the high density region.  The transition
line in the $(d,B)$ phase diagram at $T=0.10$ can be fitted closely
with the power-law $B \sim d^{0.438~(0.436)}$ for the multiquark
with $n_{s}=0~(0.2)$.  On the other hand, the transition line in the
$(d,T)$ phase diagram is weaker than the logarithmic of the density
but nevertheless it is an increasing function with respect to the
density.  These imply that for sufficiently large density, the
chirally broken multiquark phase is the most preferred nuclear phase
even in the presence of the external magnetic field.

The situation when density becomes extremely large and being
dominant occurs in the core of dense star such as the neutron star.
Therefore it is very likely that the core of dense warm star
composes primarily of the multiquark nuclear matter even when an
enormous magnetic field is present such as in the core of the
magnetars. It is possible that a large population of the warm
magnetars has multiquark cores.  These warm dense objects could be
relatively more massive than typical neutron stars.

\section*{Acknowledgments}
\indent  I would like to thank CERN Theory Division for the warm
hospitality during my visit where this work is completed.  P.B. is
supported in part by the Thailand Research Fund~(TRF) and Commission
on Higher Education~(CHE) under grant RMU5380048 and Thailand Center
of Excellence in Physics~(ThEP).

\appendix
\section{Force condition of the multiquark configuration}

Fixing the characteristic scale $L_{0}$ to $1$ for the brane
configuration requires balancing three forces in the gravity
picture.  The D8-brane tension must be in equilibrium with the tidal
weight of the D4 source and the string tension of the colour
strings.  The derivation of the $x'_{4}(u_{c})$ presented here is
the same as in Ref.~\cite{pbm}, it is included for completeness.

We vary the total action with respect to $u_{c}$ to obtain the
surface term. Imposing the scale-fixing condition
$2\int^{\infty}_{u_{c}}~du x^{\prime}_{4}(u)=L_{0}=1$, we found
that~\cite{bll}
\begin{eqnarray}
x^{\prime}_{4}(u_{c})& = & \displaystyle{ \left( \tilde{L}(u_{c})
-
\frac{\partial{S_{source}}}{\partial{u_{c}}}\right)\Bigg{/}{\frac{\partial
\tilde{S}}{\partial{x^{\prime}_{4}}}\bigg{\vert}_{u_{c}}}},
\label{app1}
\end{eqnarray}
as the condition on $u_{c}$.

We perform the Legendre transformed action with respect to
$a^{V\prime}_{0}$ and $a^{A\prime}_{1}$ to obtain
\begin{eqnarray}
\tilde{S}& = &
\int^{\infty}_{u_{c}}\tilde{L}(x^{\prime}_{4}(u),d)\,du, \nonumber \\
              & = &
\mathcal{N}
\int^{\infty}_{u_{c}}du~\sqrt{\frac{1}{f(u)}+u^{3}x_{4}^{\prime
2}} \nonumber
\\
& \times & \sqrt{f(u)(C(u)+D(u)^{2})-\Big(j_{A}-\frac{3}{2}B\mu
+3Ba_{0}^{V}\Big)^{2}},\label{app2}
\end{eqnarray}
where $C(u)\equiv u^{5}+B^{2}u^{2},D(u)\equiv d+3Ba_{1}^{A}(u)-3B
\mathcal{5}\varphi/2$.  Note that the Chern-Simon action are
included in the total action during the transformations.

The Chern-Simon term with the derivatives $a^{V \prime},a^{A
\prime}$ eliminated is
\begin{eqnarray}
S_{CS}& = & -\mathcal{N} \frac{3}{2}B
\int^{\infty}_{u_{c}}du~\frac{\Big(
a^{V}_{0}(j_{A}-\frac{3}{2}B\mu +3Ba_{0}^{V})-f(u)
D(u)a^{A}_{1}\Big)\sqrt{\frac{1}{f(u)}+u^{3}x_{4}^{\prime
2}}}{\sqrt{f(u)(C(u)+D(u)^{2})-\Big(j_{A}-\frac{3}{2}B\mu
+3Ba_{0}^{V}\Big)^{2}}}.\label{app3}
\end{eqnarray}

From Eqn.~(\ref{app1}),(\ref{app2}),(\ref{app3}),(\ref{app4}), and
the boundary conditions,
$a^{V}_{0}(u_{c})=\mu_{source},a^{A}_{1}(u_{c})=0$, we can solve to
obtain the condition for the static equilibrium
\begin{eqnarray}
(x^{\prime}_{4}(u_{c}))^{2}& = &
\frac{1}{f_{c}u_{c}^{3}}\Big[\frac{9}{d^{2}}\frac{
(f_{c}(C_{c}+D_{c}^{2})-(j_{A}-\frac{3}{2}B\mu+3 B
a_{0}^{V}(u_{c}))^{2})}{(1+\frac{1}{2}(\frac{u_{T}}{u_{c}})^{3}+3
 n_{s}\sqrt{f_{c}})^{2}}
-1\Big], \nonumber
\end{eqnarray}
where $f_{c}\equiv f(u_{c}),C_{c}\equiv C(u_{c}),D_{c}\equiv
D(u_{c})$.

\newpage

\end{document}